\documentclass[journal=jpca,manuscript=article]{achemso}

\usepackage[version=3]{mhchem} 
\usepackage{gensymb}
\usepackage{xcolor}
\usepackage{comment}
\usepackage[normalem]{ulem}

\author{Eric A. Haugen}
\affiliation[UCB]
{Department of Chemistry, University of California, Berkeley, CA 94720, USA}
\alsoaffiliation[LBL]
{Chemical Sciences Division, Lawrence Berkeley National Laboratory, Berkeley, CA 94720,USA}

\author{Diptarka Hait}
\alsoaffiliation[LBL]
{Chemical Sciences Division, Lawrence Berkeley National Laboratory, Berkeley, CA 94720,USA}
\author{Valeriu Scutelnic}

\affiliation[UCB]
{Department of Chemistry, University of California, Berkeley, CA 94720, USA}
\alsoaffiliation[LBL]
{Chemical Sciences Division, Lawrence Berkeley National Laboratory, Berkeley, CA 94720,USA}
\author{Tian Xue}
\affiliation[UCB]
{Department of Chemistry, University of California, Berkeley, CA 94720, USA}
\alsoaffiliation[LBL]
{Chemical Sciences Division, Lawrence Berkeley National Laboratory, Berkeley, CA 94720,USA}
\author{Martin Head-Gordon}
\affiliation[UCB]
{Department of Chemistry, University of California, Berkeley, CA 94720, USA}
\alsoaffiliation[LBL]
{Chemical Sciences Division, Lawrence Berkeley National Laboratory, Berkeley, CA 94720,USA}
\author{Stephen R. Leone}
\affiliation[UCB]
{Department of Chemistry, University of California, Berkeley, CA 94720, USA}
\alsoaffiliation[LBL]
{Chemical Sciences Division, Lawrence Berkeley National Laboratory, Berkeley, CA 94720,USA}

\altaffiliation{Department of Physics,  University of California, Berkeley, CA 94720, USA}
\email{srl@berkeley.edu}

\title[HfAcAc Tr-XAS]
 {
 Ultrafast X-ray Spectroscopy of Intersystem Crossing in Hexafluoroacetylacetone: Chromophore Photophysics and Spectral Changes in the Face of Electron Withdrawing Groups}

\abbreviations{IR,NMR,UV}

\begin{document}

\begin{abstract}
Intersystem crossings between singlet and triplet states represent a crucial relaxation pathway in photochemical processes. Herein, we probe the intersystem crossing in hexafluoro-acetylacetone with  ultrafast X-ray transient absorption spectroscopy at the carbon K-edge. We observe the excited state dynamics following excitation with 266 nm UV light to the \textsuperscript{1}$\pi\pi^{*}$ (S\textsubscript{2}) state with element and site-specificity using a broadband soft X-ray pulse produced by high harmonic generation. These results are compared to X-ray spectra computed from orbital optimized density functional theory methods. It is found that the electron withdrawing fluorine atoms decongest the X-ray absorption spectrum by enhancing separation between features originating from different carbon atoms. This facilitates the elucidation of structural and electronic dynamics at the chromophore. The evolution of the core-to-valence resonances at the carbon K-edge reveals an ultrafast population transfer between the \textsuperscript{1}$n\pi^{*}$ (S\textsubscript{1}) and \textsuperscript{3}$\pi\pi^{*}$ (T\textsubscript{1}) states on a $1.6\pm0.4$ ps timescale, which is 
similar to the 1.5 ps timescale earlier observed for acetylacetone [J. Am. Chem. Soc. \textbf{139}, 16576 (2017)]. It therefore appears that terminal fluorination has little influence on the intersystem crossing rate of the acetylacetone chromophore. In addition, the significant role of hydrogen-bond opened and twisted rotational isomers is elucidated in the excited state dynamics by comparison of the experimental transient X-ray spectra with theory.
\end{abstract}

\section{Introduction}
Light induced chemical reactions possess the ability to redistribute energy over various electronic states and initiate changes in molecular geometry. This redistribution of energy plays a key role in a wide variety of molecular systems ranging from biological, atmospheric/interstellar, molecular switches and many more \cite{Kukura1006,Pederzoli2011,Mar1972,Maeda2015}. Fundamental understanding of the relationship between nuclear and electronic motion continues to be a primary driving force for scientific studies that seek to describe photochemistry on an ultrafast timescale. 

The goal of creating a “molecular movie”, which provides a clear step-by-step picture of both the electronic and nuclear dynamics on the timescales of nuclear motion, remains attractive in the field of physical chemistry. In order to pursue this goal, numerous techniques and methods have been employed, including photoelectron spectroscopy, X-ray absorption spectroscopy, electron diffraction and many other methods, all aiding in providing a clearer picture of the dynamics following photoexcitation.\cite{Stolow2004,Siwick1382,Wolf2017} As the fundamental understanding of standard systems grows, there will be increased demand to selectively modify molecular systems in order to fine-tune various chemical properties for use in future devices or applications. Of particular interest is acetylacetone (AcAc), which is a prototypical molecule that has been studied in depth to elucidate the ultrafast intersystem crossing between singlet and triplet states. \cite{Squibb2018,Kotsina,bhattacherjee_ultrafast_2017} The AcAc structure has many possible industrial, medical and environmental uses, as it can act as a chelating agent or serve as a structural component for UV-chromophores. \cite{KAWAGUCHI198651,Shatth,Zhang_AcAc_industrial}

\begin{figure}[h]
\centering
  \includegraphics[height=4cm]{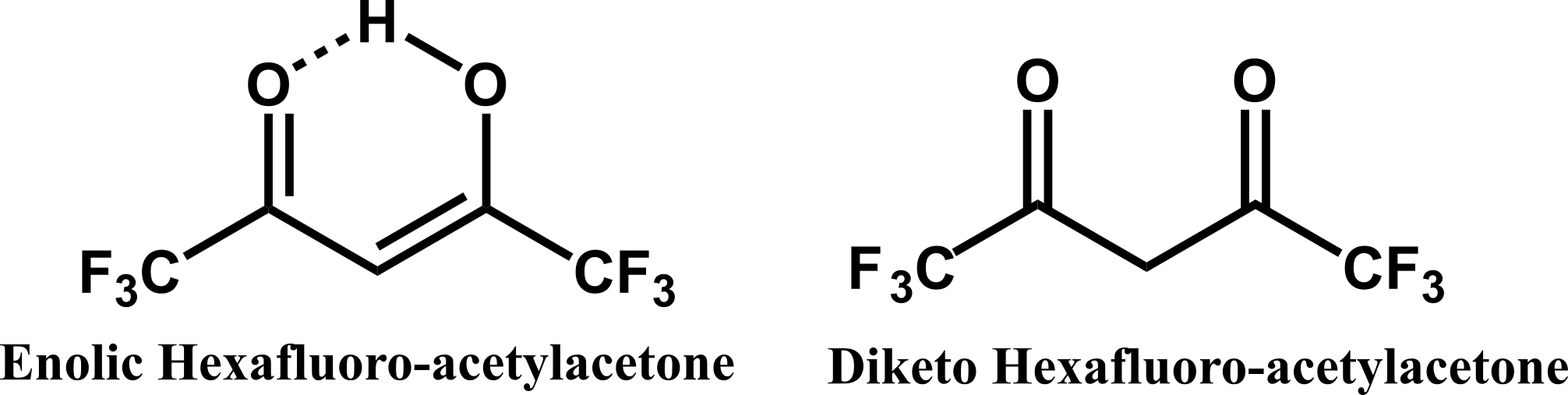}
  \caption{Structures of the enolic and diketo tautomers of hexafluoro-acetylacetone. The enolic form is more stable and is present as the dominant tautomer under experimental conditions.}
  \label{fig:hfacacstruct}
\end{figure}

Controlling the rate of the transition between the singlet and triplet states can be used to influence the quantum efficiency of various relaxation pathways, ultimately favoring certain channels over others. One potentially significant modification to AcAc is the substitution of the CH\textsubscript{3} groups with CF\textsubscript{3} groups, as shown in Fig. \ref{fig:hfacacstruct}. 

The CF\textsubscript{3} functional groups are strongly electron withdrawing and have been shown to exert a large influence on molecules, resulting in markedly different photoproducts and chemical properties\cite{Bassett1976,Yoon2000,Chen2006,Muyskens2012}. However, the fluorine substituted hexafluoro-acetylacetone (HfAcAc) has, to our knowledge, not previously been studied in the ultrafast regime. It is thus unknown to what extent the terminal CF\textsubscript{3} groups might influence the electronic state dynamics of the main chromophore of the molecule. Notably, in some conjugated chromophores it has been shown that the addition of fluorine atoms has the ability to influence the energies of the various electronic states of the molecule, thereby affecting the rate of intersystem crossing. \cite{Reiffers,bracker_impact_2019} The ability to fine-tune quantum yields of various processes is attractive for optical imaging applications or other processes such as maximizing singlet oxygen generation. Additionally, greater understanding of the mechanisms underlying intersystem crossings can prove useful for applications such as thermally activated delayed fluorescence organic light emitting diodes, where great care is taken to fine-tune the conversion between singlet and triplet states.\cite{Uoyama2012,Zhang2014}
Due to these considerations, there is an ever-increasing importance to fundamentally understand how modifying functional groups influences the various pathways corresponding to relaxation of excited electronic states.

The photodynamics of HfAcAc can be contrasted with AcAc; here the six additional fluorines act as electron withdrawing groups, heavily favoring the enol hydrogen-bonded ``ring" form of the well known keto-enol equilibrium in $\beta$-diketones\cite{Muyskens2012}. However, the strong intramolecular hydrogen bond is believed to be significantly weakened by the terminal CF\textsubscript{3} groups\cite{Chatterjee2010,DEVRIES201615}. Indeed, we computationally find that the hydrogen bond in HfAcAc is weaker by 5 kcal/mol than AcAc (as detailed in the supporting information). It is therefore easier to break the hydrogen bond in HfAcAc, thereby permitting rotation of the O-H bond about the enol C-O bond, potentially allowing different excited state dynamics to occur following electronic excitation, in this case with 266 nm light. 

\begin{table}[htb!]
\color{black}{\begin{tabular}{llll}
 State & Character  & HfAcAc & AcAc   \\
\ce{T1} & $^3\pi\pi^*$ & 2.93  & 3.31 \\
\ce{T2} & $^3n \pi^*$ & 3.73 & 4.04 \\
\ce{S1} & $^1n \pi^*$ & 4.21 & 4.46 \\
\ce{S2} &$^1\pi\pi^*$ & 5.04 & 5.21 
\end{tabular}}
\caption{Vertical electronic excitation energies (in eV) at the ground state minimum geometry for HfAcAc and AcAc, as computed from TDDFT with $\omega$B97X-D/aug-pcseg-1. $n$ refers to the lone pair on the keto oxygen.}
\label{tab:FCenergies}
\end{table}

The UV absorption spectra of both AcAc and HfAcAc share many similarities, both possessing strong and broad $\pi \rightarrow \pi ^\ast$ transitions around 266 nm corresponding to excitation into the S\textsubscript{2} state.\cite{Nakanishi1977,Nakanishi1978} Indeed, calculations reveal that the four lowest excited states have similar energies and character in the Franck-Condon (FC) region, as shown in Table \ref{tab:FCenergies}. It is therefore expected that both species will ultimately undergo intersystem crossing into the T\textsubscript{1} state following excitation to \ce{S2}, before relaxing to the ground state or dissociating. The possible role of the \ce{T2} state is not known and is not explored here. The additional fluorines allow for a greater variety of potential photoproducts in HfAcAc, permitting the production of pentafluoromethyl-3-furanone together with HF in the low pressure regime.\cite{Bassett1976,Antonov2019,KUSABA20071447} The proposed pathway to this photoproduct relies on the cleavage of the intramolecular O-H hydrogen bond,
thereby allowing the molecule to rotate and form a series of rotationally distinct isomers. These rotational isomers can act as precursors to the aforementioned photoproducts. Observation of these dynamics on ultrafast timescales is desirable to define the initial steps that the primary chromophore undergoes. In particular transient X-ray absorption spectroscopy will allow for the clear demarcation of the various singlet and triplet states from one another and can detect any potential photoproducts that might arise from the parent molecule.\cite{bhattacherjee_ultrafast_2017,Yang2018}

The field of transient X-ray absorption spectroscopy (Tr-XAS) with femtosecond timescales has made steady progress in recent years; tabletop sources producing soft X-rays via high harmonic generation have improved allowing experiments in the water window to be performed.\cite{scutelnic2021,Fu2020,Pertot264,Ross_CCl4} Tr-XAS is a powerful technique that acts as a probe for both the electronic and structural dynamics of molecules. The X-rays provide unique element specificity due to the transitions originating from localized core orbitals, and the energies corresponding to these transitions for each element vary between atoms in different locations in the molecule, resulting in spectra with site specificity. Thus core-level transitions are sensitive to the local chemical environment and report on shifts in electron density in the proximity of the probed atoms.\cite{Loh2013} Here, we use femtosecond soft X-ray transient absorption spectroscopy to probe the ultrafast non-adiabatic population transfer into the triplet $\textsuperscript{3}\pi \pi ^\ast$ (T\textsubscript{1}) state following the initial excitation of ground state HfAcAc to the optically bright $\textsuperscript{1} \pi \pi ^\ast$ (S\textsubscript{2}) state,  thereby gaining insight into the photochemical reaction pathways.

We compare the results to previous measurements on AcAc and computed spectra obtained via orbital optimized density functional theory methods.\cite{hait2021orbital} The results show that the rate of the non-adiabatic passage to the T\textsubscript{1} state in HfAcAc is not significantly influenced by the terminal fluorine atoms while providing clear evidence for hydrogen-bond ring opened and twisted rotational conformers in the excited state dynamics, in contrast to the initial hydrogen bonded, planar, enol form in the ground state.

\section{Methods}
\subsection{Experimental}
HfAcAc (98\% purity) purchased from Sigma Aldrich was stored in glass vials under refrigeration. No further purification of the samples was undertaken. The sample container was submerged in an ice-salt bath at approximately -5$\degree$ C. The HfAcAc was probed using a gas cell with a path length of 4 mm and an approximate pressure of $\approx 15$ mbar. This gas cell was further heated to 60 $^\circ$C to reduce clogging.

A comprehensive description of the experimental setup may be found in previous papers.\cite{bhattacherjee_ultrafast_2017,Attar_2017,scutelnic2021,Epshtein2020}. Briefly, a commercial Ti:Sapphire laser (Spit-fire Ace, Spectra Physics) provides a 12mJ, 1kHz, sub-45fs pulse duration output. This pulse was split by a 90:10 beamsplitter where 10\% is used to generate the UV pump and 90\% pumps an optical parametric amplifier (HE-TOPAS, Light Conversion, 1180-2600 nm). The optical parametric amplifier was tuned to produce 1470 nm (2.5-3.0 mJ/pulse) with a duration of $\approx 50$ fs. The 1470 nm pulse was used to generate a broadband soft X-ray pulse extending to the carbon K-edge (160-350 eV, sub 50 fs), which will act as the probe. The soft X-rays were focused with a gold toroidal mirror onto the sample and subsequently dispersed with an XUV variable line space grating with 1200 lines/mm (Hitachi grating 001-0660) onto a translatable X-ray camera (Princeton Instruments, PIXIS:XO 400B). The ultraviolet pump was generated from the remaining 10\% output from the Ti:Sapphire laser in a two-step process consisting of second harmonic generation resulting in production of 400 nm and sum frequency generation (SFG) of 800 and 400 nm resulting in the generation of a sub 70 fs, 266 nm pulse with a maximum energy of 120 $\mu J$/pulse, this pulse energy was detuned by rotating the SFG BBO to limit the energy to no more than 40 $\mu J$/pulse. The HfAcAc molecules were excited with the 266 nm pulses with up to $\sim$ 30 mJ/cm\textsuperscript{2} pump fluence (2.5$\times$10\textsuperscript{11} W/cm\textsuperscript{2} intensity). The molecular dynamics were then probed with the broadband, soft X-ray pulse optimized for flux at the carbon K-edge at varying time delays between the pump and probe pulses.

The time delays utilized in this experiment were increased by increments of 100 fs between time zero and 500 fs, 250 fs between 500 fs and 4 ps, and 2.5 ps between 4 ps and 20 ps. A final delay point of 30 ps is also measured. Recent experiments indicate an instrument response function of $\approx 80$ fs.\cite{scutelnic2021} This is determined by an \emph{in situ} cross-correlation of the pump and probe pulses through measurement of the ponderomotive shift of the core-excited Rydberg states of argon. The X-ray spectral resolution of the experiment is determined to be $320 \pm 30$  meV through measurement of the Gaussian broadening of the 2p\textsubscript{$3/2$} $\xrightarrow{}$ 4s transition in argon, assuming a core-hole lifetime broadening of 114 meV\cite{Sairanen_1996} (Fig.  S1).

\subsection{Theoretical}
All calculations are performed with a development version of the Q-Chem 5 package.\cite{epifanovsky2021software} C K-edge spectra are simulated with excited state specific orbital-optimized DFT (OO-DFT\cite{hait2021orbital}), utilizing the SCAN functional\cite{SCAN}. In particular, the spectra of closed-shell species are modeled with restricted open-shell Kohn-Sham (ROKS\cite{frank1998molecular,kowalczyk2013excitation}), with initial guesses generated from the recently developed static exchange approximation to ROKS [ROKS(STEX)] approach.\cite{hait2022computing} The spectra of open-shell systems are computed via $\Delta$SCF\cite{bagus1965self,besley2009self}. Naive $\Delta$SCF calculations yield spin-impure results when both unpaired up and down spins are present, but approximately spin-pure results can be obtained through use of the general recoupling procedure described previously.\cite{hait_accurate_2020, hait2021orbital} ROKS and recoupled $\Delta$SCF, with the SCAN functional, have been previously found to be accurate to $\sim 0.3$ eV in predicting C K-edge excitation energies against experiment\cite{Hait2020ROKS,hait_accurate_2020}, without any need for empirical translation of spectra. OO for electronic excited states are performed with the square gradient minimization (SGM\cite{Hait_SGM}) and initial maximum overlap method (IMOM\cite{barca2018simple}) algorithms, for restricted open-shell and unrestricted calculations, respectively. A mixed basis-set of aug-pcX-2 \cite{ambroise2018probing} at the site of the core-excitation, and aug-pcseg-1\cite{jensen2014unifying} on all other atoms is utilized  for C K-edge calculations.\cite{Hait_SGM,hait_accurate_2020,Hait2020ROKS}

Geometries are optimized at the $\omega$B97X-D\cite{wB97XD}/aug-pcseg-1 level, with this functional being chosen as it yields good performance for excited states\cite{liang2022revisiting}. The \ce{S0} and \ce{T1} states are optimized with standard ground state DFT within the singlet and triplet subspaces, respectively. The \ce{S1} state is optimized with time-dependent DFT (TDDFT), without the Tamm-Dancoff approximation (TDA).\cite{hirata1999time} The minimum energy crossing point (MECP) between the \ce{S2} and \ce{S1} states is determined at the same level of theory. TDDFT is not employed for the \ce{T1} state, as the equilibrium structures of some rotamers in this state involve breaking of a $\pi$ bond, which TDDFT (or TDA) is incapable of modeling.\cite{hait2019beyond}

\section{Results and Discussion}
\subsection{Ground State}
\begin{figure}[h]
\centering
  \includegraphics[height=8cm]{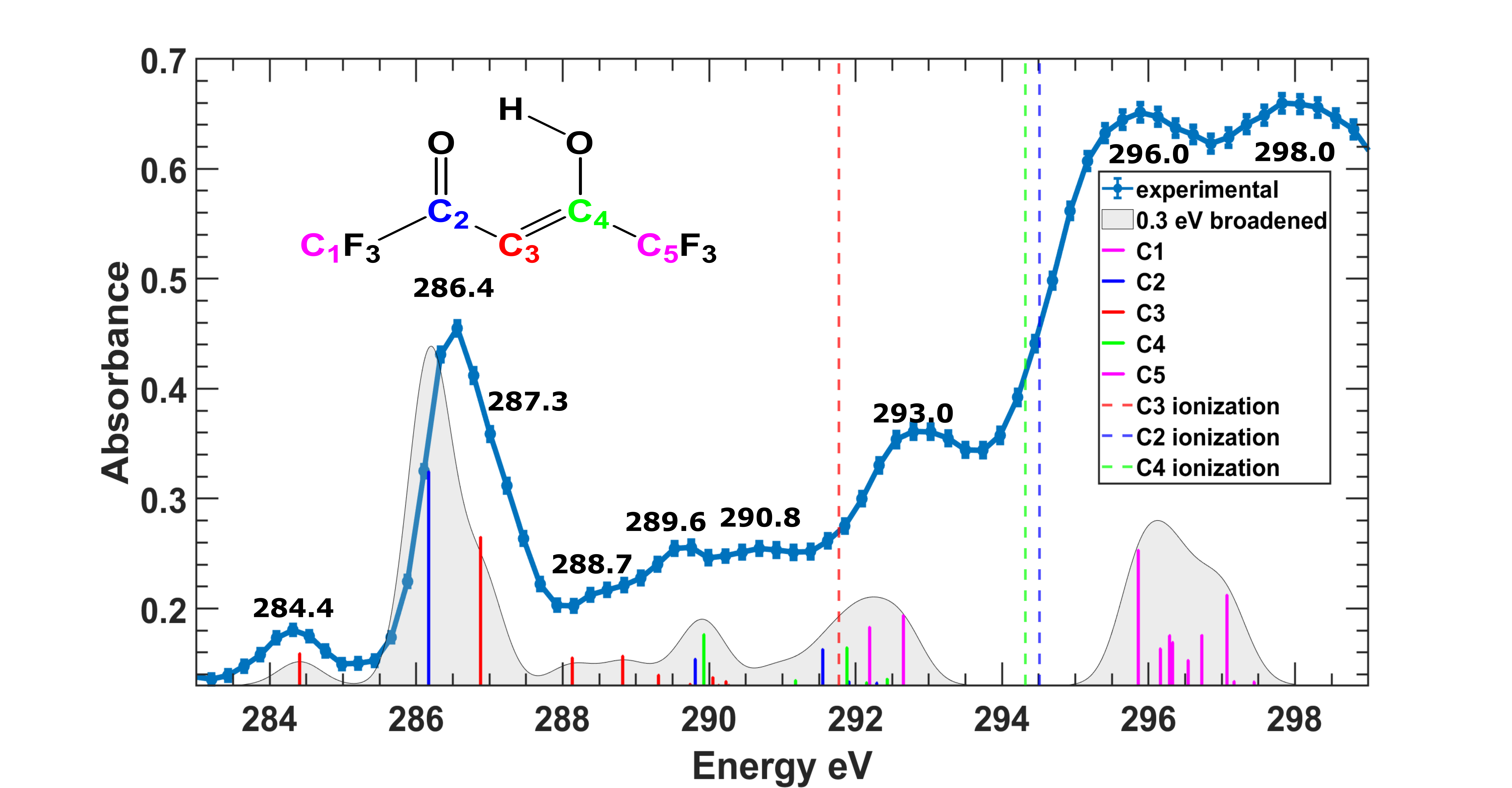}
  \caption{Experimentally measured, near-edge X-ray absorption fine-structure (NEXAFS) spectrum of HfAcAc (S\textsubscript{0}) at the carbon K-edge (solid blue line, error bars denotes 95\% confidence interval) with the main peaks annotated. The calculated ROKS spectrum of the enol tautomer is shown as a stick spectrum, color coded for distinct carbon atoms present in the molecule. The grey shaded feature corresponds in the stick spectrum broadened with a sigma=0.3 eV width Gaussian. The intensity of the theoretical spectrum is uniformly scaled to match experiment for the strong absorption feature at 286.4 eV. The calculations do not take into account the core-1\emph{s} ionization, which increases with increasing energy. The rising edge at $\sim$ 292 eV corresponds to ionization from \ce{C3}, while the rising edge at $\sim$ 294 eV arises from ionization from \ce{C2} and \ce{C4}. The dotted lines correspond to the calculated ionization energies, the ionization energy for \ce{C1} and \ce{C5} are present at $\sim$ 299 eV.
  }
  \label{fig:static}
\end{figure}

Fig. \ref{fig:static} shows the experimental near-edge X-ray absorption fine structure (NEXAFS) spectrum of HfAcAc in the \ce{S0} (ground) state, (blue points and solid blue line with error bars). The computed ROKS C K-edge spectrum for the enol tautomer is also shown. The stick spectrum displays calculated core-valence transitions of the constituent carbon atoms underlying the peaks in the observed NEXAFS spectrum. The individual computed peaks are broadened by 0.3 eV to produce the shaded grey spectrum. 
It can be clearly seen that the transitions corresponding to distinct groups of carbon atoms are significantly shifted from one another and are most prominent in their own distinct region of the X-ray spectrum. This shift arises from a chemical shift of the binding energy of the core 1\emph{s} electrons due to their proximity to more electronegative heteroatoms. OO-DFT calculations indicate that the \ce{C3} atom has the lowest 1\emph{s} binding energy (i.e. core ionization energy) at 291.7 eV, the oxygen bound \ce{C2} and \ce{C4} atoms have 1\emph{s} binding energies of 294.3-294.5 eV, while the \ce{C1} and \ce{C5} atoms directly bound to the fluorines have 1\emph{s} binding energies of 298.7-299.1 eV.  This is reflected in the X-ray absorption spectrum for the lowest energy $1s \rightarrow \pi^\ast$(LUMO) transitions. The central C\textsubscript{3} atom has the lowest energy  transition at 284.4 eV, while \ce{C2} and \ce{C4} start absorbing around 286.4 eV.

The effect of the fluorine atoms on the C\textsubscript{2}/C\textsubscript{3}/C\textsubscript{4} $1s \rightarrow \pi^\ast$(LUMO) excitation energies appears to be nearly  negligible with a maximum of a 0.2 eV shift compared to the results of acetylacetone found by Bhattacherjee \emph{et al}\cite{bhattacherjee_ultrafast_2017}. This minor effect indicates that the presence of the electronegative heteroatoms does not significantly affect the relative spacing between the core and the unoccupied valence levels in neighboring atoms, with excitation energies being significantly affected only for the atom to which the electronegative heteroatoms are bonded to. However, a large shift of around 4.8 eV for the lowest excitation out of C\textsubscript{1} and C\textsubscript{5} is found to arise from the substitution of hydrogen with fluorine. This large blue shift greatly simplifies the spectrum and helps distinguish between transitions from the terminal and central carbon atoms. Furthermore, computationally the additional fluorine atoms increase the lowest core-ionization potential of the entire molecule (corresponding to \ce{C3}) by almost 1.6 eV compared to what is observed experimentally in AcAc.\cite{bhattacherjee_ultrafast_2017} In HfAcAc the C\textsubscript{3} carbon is calculated to have the lowest vertical ionization potential at 291.7 eV, while the corresponding value for AcAc is 290.1 eV. As a result, transitions above 289 eV in AcAc, are unable to be resolved experimentally due to the onset of ionization of C 1\emph{s} electrons\cite{bhattacherjee_ultrafast_2017}. Similarly, the lowest vertical valence ionization energy for HfAcAc is computed to be 10.63 eV, while it is 9.05 eV for AcAc. 
With only minor shifts observed in the $1s\rightarrow\pi^\ast$(LUMO) transition for the  \ce{C2}, \ce{C3} and \ce{C4} atoms, it therefore appears that the influence of the fluorine atoms is primarily to lower the energies of core and valence levels associated with the central chromophore (\ce{C2}, \ce{C3} and \ce{C4}) by roughly the same amount (1.6 eV) 
 relative to an orbital infinitely far away (i.e. the continuum). Consequently, electrons are more strongly bound against ionization in HfAcAc relative to AcAc, while the excitation energies for core-to-valence or valence-to-valence transitions are relatively unaffected (as illustrated by the core-level excitation energies discussed in the text and Table \ref{tab:FCenergies} for valence excitations).  
This greatly increases the range where core-level X-ray transitions may be clearly observed prior to the onset of the core-ionization continuum, allowing significantly more transitions to be experimentally detected.

The $1s \rightarrow \pi^\ast$(LUMO) excitations from \ce{C1} and \ce{C5} are predicted to only begin at $\sim$292.2 eV (obscured in the experimental spectrum by the \ce{C3} ionization edge), which leads to a prominent peak at 293.0 eV. We note that the \ce{C1} and \ce{C5} atoms have significant oscillator strength for these transitions only because there exist C-F $\sigma^*$ levels of the same orientation as the $\pi^\ast$(LUMO). The separation between these unoccupied levels is too large in the ground state molecule for any mixing of character, as made evident by the similar core-excited state energies for HfAcAc and AcAc for the chromophore centered excitations that were previously described. However, a core-hole centered at \ce{C1} or \ce{C5} would stabilize the local C-F $\sigma^*$ levels to a significantly greater extent than the $\pi^\ast$ centered on the chromophore, through reduced electronic repulsion. This permits mixing of C-F $\sigma^*$ character into the $\pi^\ast$ LUMO for core-level X-ray transitions out of \ce{C1} and \ce{C5}, leading to some oscillator strength for the 1\emph{s}$\to \pi^\ast$ transitions. This behavior is therefore different from the stabilization of core and valence chromophore orbitals by core-hole free \ce{CF3} moieties that were described earlier.

The shoulder in the \ce{S0} spectrum at $\sim$ 287 eV can be solely assigned to the \ce{C3} 1\emph{s} $\to \pi^\ast$(LUMO+1) transition.  The broad region at $\sim$ 288.7 eV arises from \ce{C3} 1\emph{s} as well, corresponding to excitations to $\sigma^*$ and Rydberg levels. Higher energy peaks observed, such as at 289.6 and 290.8 eV, correspond to a mixture of overlapping transitions of the 1\emph{s} core-electrons of \ce{C2}, \ce{C3} and \ce{C4} to higher energy unoccupied valence and Rydberg orbitals. Improved experimental resolution would be needed to distinguish between the transitions and to assign them to distinct carbon atoms. Finally, the peaks at 296.0 and 298.0 eV correspond to transitions of the CF\textsubscript{3} carbons and are composed of numerous excitations to higher unoccupied valence and Rydberg orbitals.

It should be noted that previous studies on HfAcAc\cite{Muyskens2012} have demonstrated that the enolic tautomer will be the dominant configuration of HfAcAc with $ >99\%$ contribution under current experimental conditions. Due to this, the contribution of the diketone tautomer of the HfAcAc is negligible and its effect on the ground state spectrum was not considered.

As noted earlier, the substitution of CF\textsubscript{3} type carbons for CH\textsubscript{3} carbons pushes the transitions associated with those carbon atoms to higher energies and allows for the clear distinction between the terminal carbons $1s\rightarrow \pi ^\ast$ and the central carbons $1s\rightarrow \text{Rydberg}$ transitions. This allows greater distinction between the  carbon peaks of HfAcAc, allowing them to be more clearly defined and isolated from one another. This demonstrates the possibility of adding strong electron withdrawing groups to complex systems in future experiments to help isolate and deconvolute electronic transitions at the carbon K-edge.

\subsection{Excited State Dynamics}
\begin{figure*}[btp]
\centering
  \includegraphics[height=8cm]{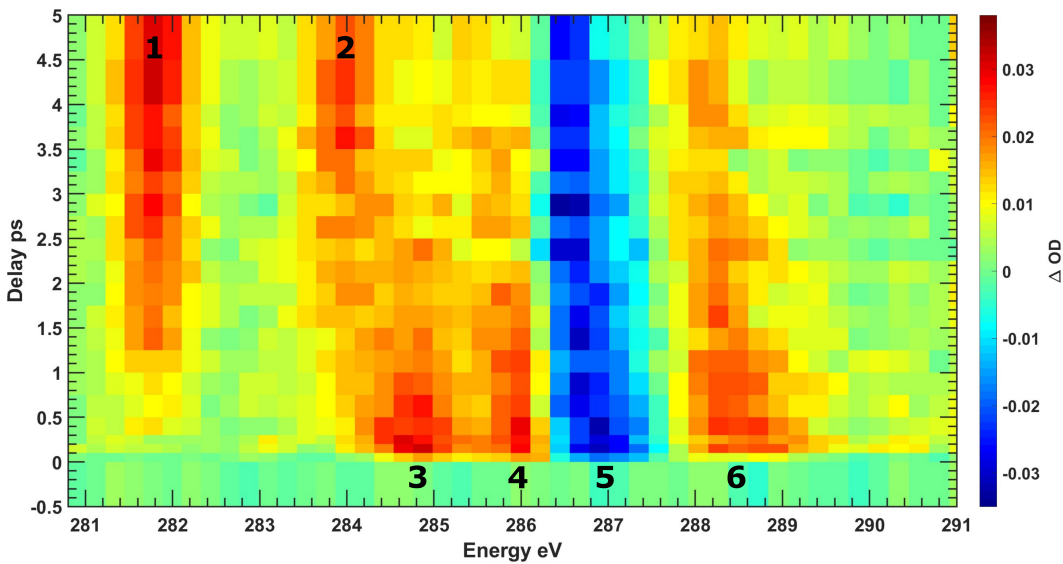}
  \caption{Experimentally measured, transient spectrum versus time of HfAcAc following excitation with 266 nm pump pulse.}
  \label{fig:colourmap}
\end{figure*}

Fig. \ref{fig:colourmap} shows a two-dimensional false color transient differential absorption map of the transient C K-edge X-ray spectrum of HfAcAc after excitation with a 266 nm UV pulse for time delays of -500 fs to 5 ps. All experimental signals correspond to a single photon excitation, as may be seen in Fig S2. Data taken at longer time delays do not significantly deviate from the 5 ps results (as shown in Fig. S5). 
The red-yellow and blue colors represent positive and negative $\Delta$OD features, respectively. Here the pump-off spectrum was used to determine the correlation matrix that was employed for the edge-pixel referencing technique, reducing noise originating from fluctuations in the high harmonic generation process which is present in transient absorption data.\cite{Geneaux:21}

Six easily distinguishable  features are seen in Fig. 3, which are labeled 1-6. OO-DFT calculations and comparison to previous results for AcAc\cite{bhattacherjee_ultrafast_2017} indicate that several of these features can be assigned to the \textsuperscript{1}$n\pi^{*}$ (S\textsubscript{1}) and \textsuperscript{3}$\pi\pi^{*}$ (T\textsubscript{1}) states (where $n$ corresponds to the non-bonding 2p lone pair on the ketone O atom). The intersystem crossing between these states occurs in $1.6\pm 0.4$ ps, as discussed later.

OO-DFT calculations reveal that peaks 1 and 2 (281.7, 283.9 eV) correspond to the T\textsubscript{1} state , while peaks 3 and 4  (284.8, 285.9 eV) correspond to the S\textsubscript{1} state. Peak 5 corresponds to the depleted \ce{S0} state and is centered at 286.3 eV. An energy shift on the right shoulder of the ground state depletion at increasing delays indicates a transition from singlet to triplet states. Peak 6 (288.6 eV) likewise corresponds to the S\textsubscript{1} and T\textsubscript{1} states, where a clear energy shift occurs, corresponding to passage through the intersystem crossing. Higher energy transient signals seemingly unrelated to the aforementioned intersystem crossing are also present. These transitions will primarily arise from the terminal \ce{CF3} carbons, which are discussed further in the supporting information.

The transient data at energies below the primary ground state depletion shows very similar characteristics to that of AcAc. The energies of low energy peaks assigned to the triplet T\textsubscript{1} state and the singlet S\textsubscript{1} state (281.7, 283.9, 284.8, 285.9 eV, respectively) are very similar to the corresponding energies found for AcAc by Bhattacherjee \emph{et al.}\cite{bhattacherjee_ultrafast_2017} (281.4, 283.8, 284.7, 285.9 eV, respectively). These low energy features are a result of transitions originating from the central chromophore consisting of \ce{C2}, \ce{C3}, \ce{C4}, which are identical for both molecules. This is analogous to what was observed in the ground state static spectrum where the presence of the fluorines did not strongly affect core-excitation energies from \ce{C2}, \ce{C3}, \ce{C4}. As with the ground state, higher energy features can be clearly resolved owing to the blueshift of excitations/ionizations corresponding to the terminal CF\textsubscript{3} carbons. In particular, peak 6 (which is higher in energy than the ground state bleach) does not have an analogue in the unsubstituted AcAc transient spectrum.

\begin{figure}[h]
\centering
  \includegraphics[height=8cm]{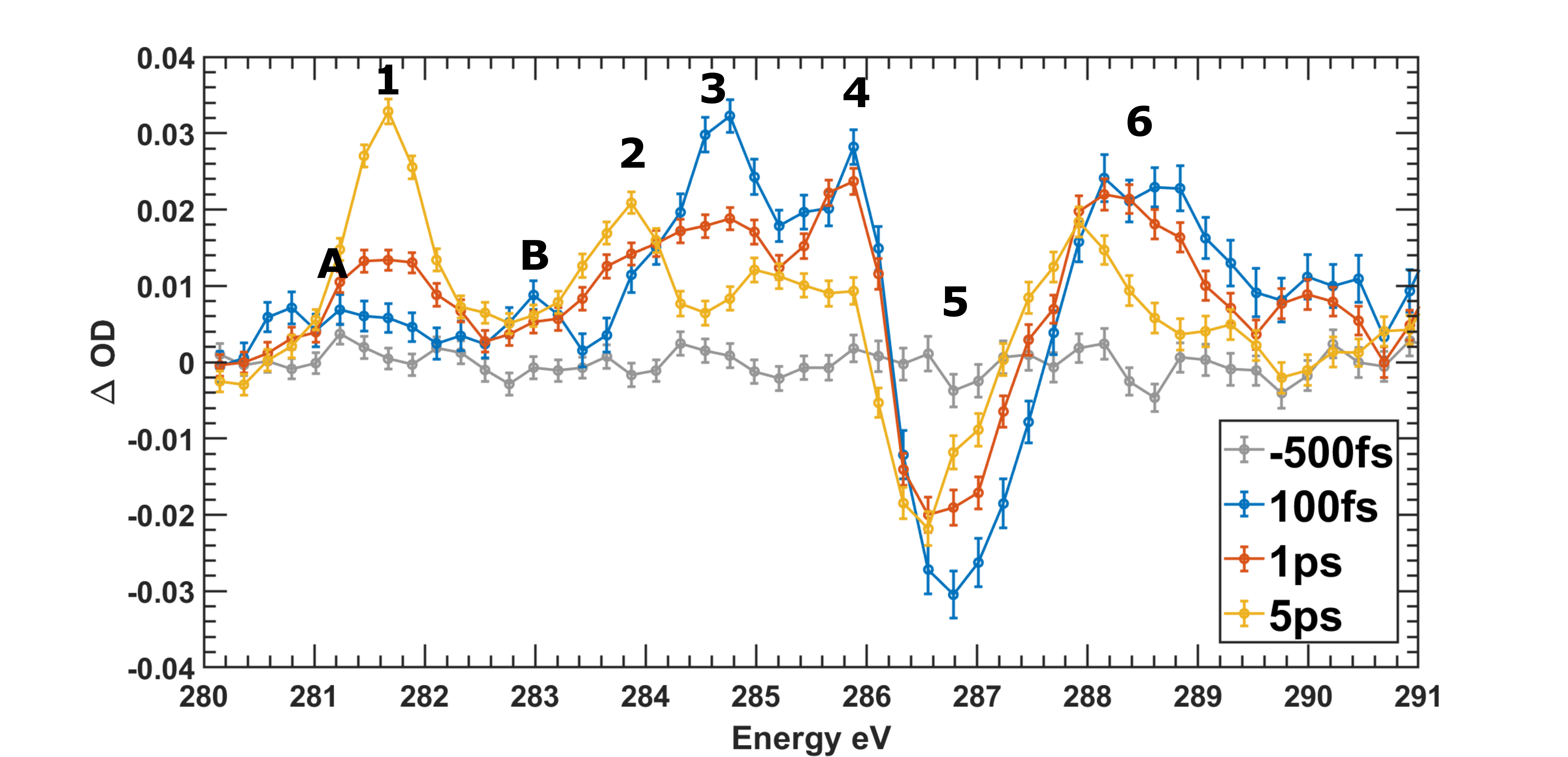}
  \caption{$\Delta$OD at 100 fs, 1 ps, and 5 ps illustrating the shift from the singlet S\textsubscript{1} state to the triplet T\textsubscript{1} state. Peaks 1-6 correspond to the transitions observed in Fig. 3, meanwhile transitions A and B correspond to weak $1s\rightarrow n$ (singly occupied oxygen 2p lone-pair) transitions consistent with the S\textsubscript{1} state, on the basis of OO-DFT calculations. }
  \label{fig:lineouts}
\end{figure}
Fig. \ref{fig:lineouts} shows experimental transient XAS spectra at selected time delays that convey the clear energy shift due to the \ce{S1\to T1} intersystem crossing. Here 100 fs represents primarily \ce{S1} character, 1 ps represents a mixture of \ce{S1} and \ce{T1}, and 5 ps represents primarily \ce{T1} character. The -500 fs timepoint is also presented as a reference for the intrinsic noise of the experiment, as it should ideally have $\Delta\text{OD}=0$ throughout. In this figure a clear transition between the features of the singlet and triplet can be distinguished. All time delays longer than 5 ps did not reveal any further significant changes in the spectrum, indicating that the triplet state is long-lived up to at least 30 ps and that photodissociation of HfAcAc occurs on a longer timescale. Longer time delays are observed in the supplement in Fig. S5. In addition to features 1-6, weak peaks at 281.2 and 283.0 eV in the 100 fs trace are labeled as features A and B, respectively. These weak peaks correspond to excitations to the singly occupied oxygen 2p lone-pair (n) orbital in the \ce{S1} state. The aforementioned assignments will be justified below through the comparison of the experimental results with the theoretical calculations.

\begin{figure}[h]
\centering
  \includegraphics[height=8cm]{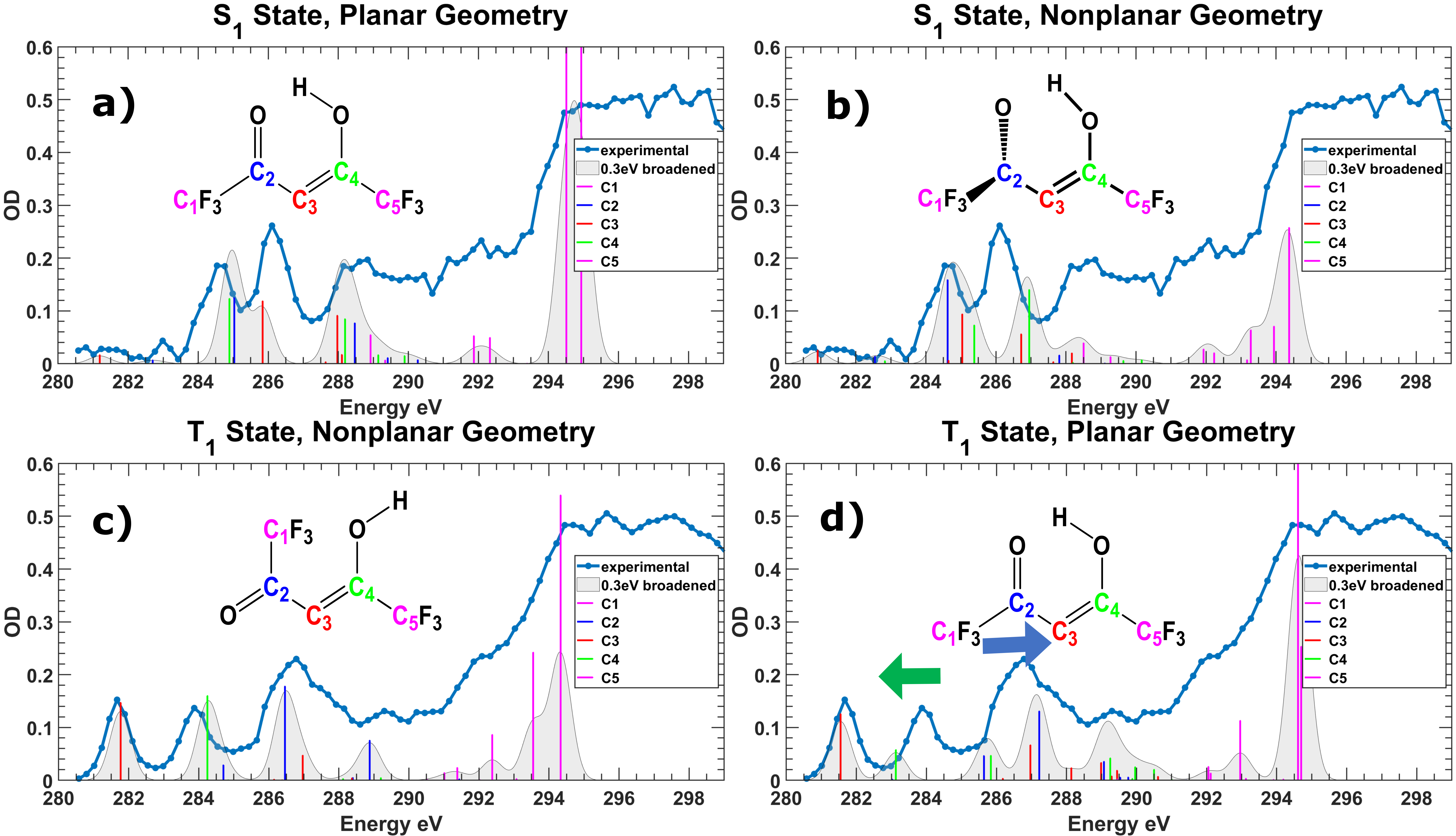}
  \caption{Experimentally determined S\textsubscript{1} and T\textsubscript{1} states compared with OO-DFT calculations. The timepoint of 100 fs was chosen as the experimental S\textsubscript{1} state and timepoints from 5-30 ps were averaged for the experimental T\textsubscript{1} state. a) S\textsubscript{1} electronic state at its optimized geometry, where the $\pi$ system (\ce{O=C2C3C4}) is planar (0.0$\degree$ dihedral). b) S\textsubscript{1} electronic state at the geometry of the S\textsubscript{2}/S\textsubscript{1} MECP geometry where a 34.4$\degree$ dihedral angle is present in $\pi$ system (\ce{O=C2C3C4}). c) \ce{T1} electronic state for rotational isomer with a broken hydrogen bond, which permits free rotation.  d) T\textsubscript{1} electronic state with a planar structure with the intramolecular hydrogen bond still intact. Arrows are included in d) to highlight predicted energy shifts for the calculated low-energy \ce{C2}/\ce{C4} transitions, driven by changes from non-planar to planar geometry. }
  \label{fig:exstate}
\end{figure}

Fig.  \ref{fig:exstate} illustrates the theoretical excited state spectra for the \ce{S1} state at two geometries and two rotational isomers (rotamers) of the \ce{T1} state, with the experimental spectrum for both states taken at two different time delays. In Fig. \ref{fig:exstate}, the different color bars correspond to the different carbon atoms and the shaded grey spectrum corresponds to a 0.3 eV broadened spectrum. The blue traces correspond to the excited state spectrum obtained via the addition of the static \ce{S0} spectrum to the transient resolved spectra, assuming a 20\% excitation from the ground state (which is the minimum fraction required to avoid any unphysical negative absorbance in the excited state spectrum, which may arise from an improper subtraction the ground state signal).
The geometries corresponding to these excited states are provided in the supporting information. Additional computed XAS for the S\textsubscript{2}, S\textsubscript{1}, and potential photoproducts are also shown in Fig. S6. 

Briefly, the calculated $^1\pi \pi^\ast$ (S\textsubscript{2}) state spectrum in the unrelaxed (Franck-Condon) geometry is predicted to have relatively strong ``fingerprint-like" features at 280.3 and 282.1 eV arising from transitions to the singly occupied $\pi$ bonding (HOMO-1) level, which are not observed in the experimental data. Indeed, no obvious rapid changes in $\Delta$OD, indicative of changing the electronic state, are observed in short time scans between 0 and 100 fs (as depicted in Fig. S7). 
This indicates that the internal conversion between the S\textsubscript{2}/S\textsubscript{1} states occurs on a timescale faster than the temporal resolution of the experiment ($\sim$80 fs). Likewise, 
it is unknown whether the intersystem crossing follows El-Sayed's rules (corresponding to a direct \ce{S1}$\to$\ce{T1} transition) or if the T$_2$ state acts as a very short-lived intermediate form
\cite{bhattacherjee_ultrafast_2017,Squibb2018}. Nevertheless, it is expected that the lifetime of the T\textsubscript{2} state would be too short to build up any significant population before relaxing to the T\textsubscript{1} state, thus rendering it difficult to experimentally observe it with our setup.

Fig.  \ref{fig:exstate}a) demonstrates the comparison between the experimental data at 100 fs and the OO-DFT computed spectra of the S\textsubscript{1} state at the optimized global minimum, planar geometry for this state. Clear discernible features at 281.2, 283.0, 284.6, 286.1 and 288.5 eV are observed in the experimental data, with higher energy features also present. The OO-DFT calculated spectrum predicts the first two, relatively weak, peaks of the spectrum, which arise from 1\emph{s}$\to n$ transitions from \ce{C3} and C$_{\{2,4\}}$ respectively. However, the theory underestimates the strength of the second peak and slightly shifts the predicted energy of the second peak to 282.7 eV.  The strong experimental peak at 284.6 eV can be assigned to calculated 1\emph{s}$\to\pi^\ast$ transitions from \ce{C2} and \ce{C4} at 285.0 eV and 284.9 eV. A relatively strong excitation from \ce{C3} is predicted at 285.8 eV. The corresponding experimental feature is at 286.1 eV, which is perceptibly more intense than what is predicted by theory. The higher energy feature at 288.5 eV is predicted by the OO-DFT calculations to correspond to other transitions from central carbon atoms.

Fig.  \ref{fig:exstate}b) compares the same 100 fs experimental data to the computed OO-DFT spectrum for the S\textsubscript{1} state at the S\textsubscript{2}/S\textsubscript{1} MECP geometry. Notably this configuration possesses a large dihedral angle of approximately 34$\degree$ for the central three carbon atoms and the carbonyl oxygen. This deviation from planarity is driven by the $\pi\pi^\ast$ character present in the S\textsubscript{2} state, which allows easy deformation of the planar $\pi$ system. The computed spectrum at this geometry shows significant differences from the spectrum computed for the S\textsubscript{1min} geometry shown in Fig \ref{fig:exstate} a), 
particularly in the range of 284.5-289 eV. The strong feature at 288.5 eV is absent, while there is a greater splitting in the predicted \ce{1s \to \pi^*} transitions, seen experimentally at 284.6 eV and 286.1 eV.

These differences arise from the nonplanarity of the molecule in this geometry, which leads to significant changes to the $\pi$ orbital system. It is worth noting that the experimental 286.1 eV signal lies between the corresponding peaks computed at the planar global minimum geometry in \ref{fig:exstate}a) and the nonplanar geometry in \ref{fig:exstate} b). This suggests that the 100 fs experimental spectrum is probably best described by an ensemble of possible molecular geometries in the \ce{S1} state. Indeed, $\sim$ 18 kcal/mol energy is available to the \ce{S1} state after relaxation from \ce{S2}, which is sufficient to explore a broad range of possible structures (the range of rotamer energies in the \ce{S1} state is 13 kcal/mol). It has also been previously reported in Tr-XAS measurements that a Wigner distribution of vibrationally excited states was needed to appropriately match calculated X-ray spectra to experimental results.\cite{scutelnic2021} It is possible that future work involving vibrationally excited trajectories may prove necessary to more accurately model the experimental \ce{S1} state spectrum. We also note that OO-DFT predicts a lower intensity for the 286.1 eV peak in the experimental spectrum [as shown in Fig \ref{fig:exstate} a) and b) ], and this discrepancy can possibly also be resolved by sampling over more geometries.

To obtain an experimental T\textsubscript{1} state spectrum for Figs. \ref{fig:exstate}c) and \ref{fig:exstate}d), the transient data from 5 to 30 ps were averaged and a portion (20\%) of the \ce{S0} state spectrum was added back under the assumption of a 20\% excitation, thus simulating a ``pure" triplet state. 
It is known that upon UV excitation the HfAcAc molecule is able to undergo rotation and form a number of rotational isomers (rotamers).\cite{NAGASHIMA200359} These rotamers involve rotations about the \ce{C2-C3}, \ce{C3=C4} and \ce{C4-OH} bonds, with each of these bonds either in ``cis" (C) or ``trans" (T) configurations.\cite{NAGASHIMA200359} The notation used in the literature consists of three letters describing the orientation for each bond.  Chemical structures displaying these orientations are shown in Fig. S9. The ground state enol form shown in Fig \ref{fig:hfacacstruct} is the triple cis or CCC rotamer, which is strongly favored energetically at the \ce{S0} state due to the hydrogen bond.  However, we find that the rotamer energies in the \ce{T1} state differ by less than 2 kcal/mol from one another, significantly less than the $\sim 44$ kcal/mol energy gained following relaxation from the S\textsubscript{2} excited state. Nonplanar configurations like the twisted CCT/TCC/TCT forms therefore are quite accessible in the \ce{T1} state. Indeed, the $^3\pi\pi^\ast$ character of the state permits relatively easy rotation about the \ce{C3-C4} bond, causing multiple non hydrogen bonded rotamers to collapse into the same, twisted, structure. For example, geometry optimization on the distinct TCT and TTT rotamers, on the \ce{T1} state, lead to the same local minimum. 
 Both the \ce{C=C} bond and the hydrogen bond are broken in such twisted structures.

Fig. \ref{fig:exstate} c) shows that the experimental spectrum for the \ce{T1} state has clear peaks at 281.7, 283.9, 286.8, 287.5 and 289.3 eV, in addition to some higher energy features. OO-DFT calculations find that all nonplanar rotamers of the \ce{T1} state are predicted to have a very similar C K-edge spectrum (as shown in Fig. S9), and thus only the computed results for the TCT state are shown in Fig. \ref{fig:exstate} c). Experiment and theory agree fairly well with each other (especially for the features at 281.7, 283.9, 286.6, 289.3 eV), indicating that a large proportion of molecules in the \ce{T1} state exist in hydrogen-bond cleaved, twisted forms.

Meanwhile, it can be seen that in Fig. \ref{fig:exstate}d) the planar CCC form of the \ce{T1} state does not adequately reproduce the experimental data on its own. Most notably, the peak at 283.9 eV has the calculated transition at 283.1 eV and is predicted to be of much weaker intensity. Agreement for the first peak at 281.7 eV however is good, and some of the higher energy features align relatively well with experiment (such as the one at 289.3 eV). Indeed, a comparison of Figs. \ref{fig:exstate}c) and \ref{fig:exstate}d) suggests that the \ce{T1} state molecules may exist as a mixture of planar and twisted nonplanar forms. The computed spectra for different ratios of these planar and nonplanar forms are shown in Fig. S10, which reveals that significant features at 283.1 and 286 eV would be displayed if the proportion of planar molecules was in excess of $\sim$35\%. These predicted transition energies are not directly overlapping with the experimentally observed peaks, thereby indicating that the majority of the molecules are likely present in a twisted form. This is not surprising, as there are a greater number of nonplanar local minima than planar local minima for the T$_1$ state of HfAcAc, with the energy difference between these states being quite low relative to the energy available to the molecule.

In this regard, it is worth noting that while peak 1 (281.7 eV, first T\textsubscript{1} transition in Fig. \ref{fig:colourmap}) is nearly identical for the planar and non-planar rotamers, peak 2 (283.9 eV, second T\textsubscript{1} transition in Fig. \ref{fig:colourmap}) is not present in the planar CCC state. This suggests the rotation must occur rapidly along the potential energy surface since there is no observed delay in the experimental data between peaks 1 and 2 within the uncertainty of the measurements. We also note that the TCC rotamer has been suggested as a precursor to the transition state for the proposed photodissociation pathway to form HF + pentafluoromethyl-3-furanone.\cite{Muyskens2012}

Despite this, the possible pentafluoromethyl-3-furanone (PF3F) product, is not observed in the experimental spectrum.  The PF3F molecule in the \ce{S0} ground state is predicted by ROKS to have a C K-edge spectrum very similar to HfAcAc ground state (as shown in Fig. S6). Therefore, formation of the \ce{S0} state PF3F would lead to a decay in the \ce{T1} signals and reduction in the ground state bleach. However, such behavior is not clearly evident in the experimental data and the transient spectrum remains virtually constant from 5-30 ps (as shown in Fig.  S5). Given the signal-to-noise ratio of the experiment, even a 15\% conversion of \ce{T1} HfAcAc into the \ce{S0} PF3F product should be readily observable.  The PF3F could potentially form in the \ce{T1} state as well, and the computed spectrum for this electronic state is very similar to the CCC form of HfAcAc in the \ce{T1} state shown in Fig. \ref{fig:exstate} d). Formation of the \ce{T1} state PF3F therefore should correspond to a depletion in peak 2 over time, as nonplanar \ce{T1} HfAcAc would be consumed to produce PF3F. No obvious decay in peak 2 is observed, indicating that a significant amount of PF3F is not forming on the timescales studied in this experiment, given the signal-to-noise ratio of the setup. Indeed, the lack of decrease in triplet intensity over the experimental timescales suggests that  the lifetime of the triplet is significantly longer than 30 ps. This indicates that the relaxation back to the ground state from the triplet is delayed and any photoproducts may occur at longer timescales.

\begin{figure}[h]
\centering
  \includegraphics[height=8cm]{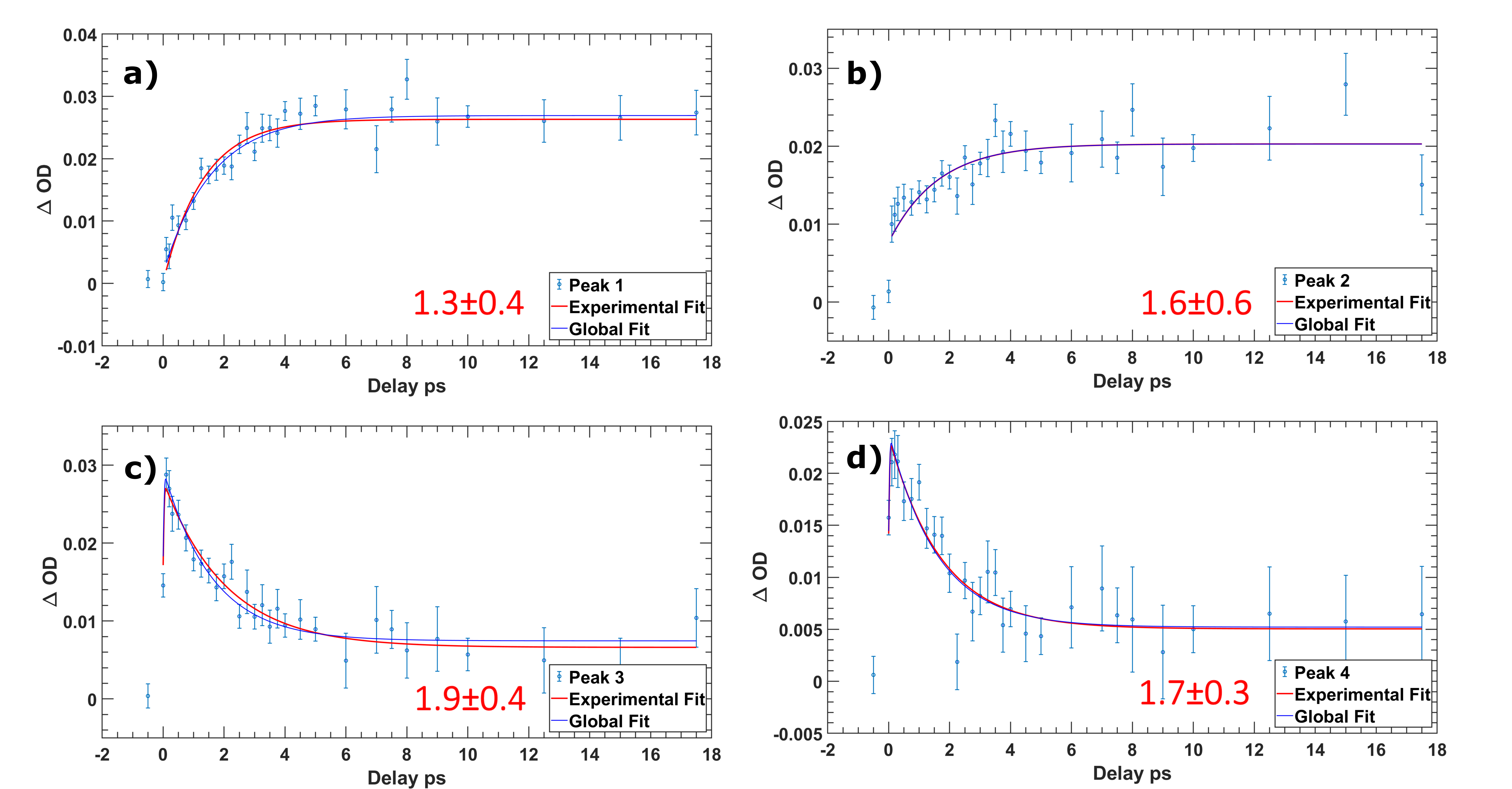}
  \caption{Fits for peaks 1-4, each consisting of an individual exponential fit corresponding to each peak and a 1.6 $\pm$ 0.4 ps exponential determined via global fitting, both are convoluted with the IRF.}
  \label{fig:temporal_lineouts}
\end{figure}

Fig.  \ref{fig:temporal_lineouts} illustrates the various time evolutions of the features in the experimental data. Each of these were fit with an exponential and a Gaussian error function to incorporate the instrument response function (IRF) of 80 fs. A global fitting routine was used to determine the fit with a single exponential for peaks 1-4. It was determined that an exponential with a time constant of 1.6 $\pm$ 0.4 ps corresponds to the intersystem crossing from the \ce{S1} state to the \ce{T1} state. 
Somewhat surprisingly there is only a minor change in the rate of the intersystem crossing when compared to AcAc, where a \ce{S1} lifetime of 1.5 ps was measured. It should be noted that the features showing the clearest signal correspond to the central three carbon atoms. These central carbon atoms are part of the $\pi$ backbone, which extends to the two oxygen atoms. In the transient data there are no clear changes in the high energy carbon peaks that occur on the timescale of the intersystem crossing. This indicates that the CF\textsubscript{3} carbons do not play a large role in this conversion from \ce{S1} to \ce{T1}, which is mostly driven by the $\pi$ system composed of the central C atoms and the O atoms. 
This raises the question whether a hypothetical electron withdrawing or donating group attached to the central C\textsubscript{3} carbon atom would produce a significant change in the rate of intersystem crossing. This substitution could have a more significant effect on the conjugated $\pi$ system present on the acetylacetone backbone. 

Additional fitting for peaks 5-6 is shown in the supporting information; the fits of peaks 5 and 6 reinforce the rate constants determined by the fit performed for peaks 1-4 and support the two step kinetic model. There are also higher energy peaks that appear to correspond to processes faster than intersystem crossing, which are discussed further in the supporting information.

\section{Conclusions}

Time-resolved X-ray absorption spectroscopy was used to investigate the non-adiabatic relaxation dynamics of hexafluoro-acetylacetone following 266 nm excitation. It was found that the intersystem crossing between the \ce{S1\to T1} occurs in $1.6\pm0.4$ ps. OO-DFT calculations  are used to assign features of the C K-edge spectra of the \ce{S0}, \ce{S1} and \ce{T1} states, and also aid in determining that hydrogen bond cleavage occurs prior to/immediately following the intersystem crossing from the singlet S\textsubscript{1} to the T\textsubscript{1} state. The \ce{T1} state is therefore present as a mixture of rotational isomers (rotamers). 
Additionally, it was determined that even the S\textsubscript{1} state is present over a range of geometries, with OO-DFT calculations illustrating the effect of the dihedral angle on the predicted positions of the $1s \rightarrow \pi^\ast$ transitions.

It was observed that the substitution of multiple fluorine atoms does not significantly impact the intersystem crossing rate between the singlet S\textsubscript{1} and triplet T\textsubscript{1} states in HfAcAc compared to AcAc. However the addition of these fluorine atoms significantly blueshifts the ionization continuum of the carbon K-edge, emphasizing higher energy features in the time-resolved spectra and allowing more possible states to be observed. The large shifts observed through the different chemical environments surrounding the various carbon atoms opens the doorway to studying different fluorinated species where these shifts may simplify the XAS spectra of other prototypical molecules.

The dynamics relating to crossing between the S\textsubscript{2} and S\textsubscript{1} states could not be resolved in this experiment due to a combination of limited experimental temporal resolution and similarity between the expected spectra of the S\textsubscript{2} and S\textsubscript{1} states. In 
the future, experiments at the oxygen K-edge may provide more insights regarding this transition; this is because there are larger differences in spectra expected in the oxygen region of the singlet excited states.\cite{Faber2019}

We demonstrate that time-resolved X-ray absorption spectroscopy using a broadband carbon K-edge high-harmonic probe reveals an ultrafast intersystem crossing between the S\textsubscript{1} and T\textsubscript{1} states in HfAcAc. Elucidation of the roles played by \ce{S2} and \ce{T2} states remain to be identified by measurements at a faster time resolution. Overall, these results further reinforce the applicability of time-resolved X-ray absorption spectroscopy and its ability to act as a powerful probe to elucidate non-adiabatic dynamics in photoexcited molecules.

\section*{Conflicts of interest}
The authors declare the following competing interest: M.H.-G is a part-owner of Q-Chem, which is the software platform in which the quantum chemical calculations were carried out.

\section{Supporting Information}
The supporting information provides further details regarding the experimental methods, power dependence studies, experimental results regarding high energy channels, shorter and longer time dynamics than shown in the primary publication, additional theoretical calculations regarding possible photoproducts and electronic states of interest.

\begin{acknowledgement}

This research was supported by Director, Office of Science, Office of Basic Energy Sciences, of the U.S. Department of Energy under Contract No. DE-AC02-05CH11231, through the Gas Phase Chemical Physics program (E.A.H, V.S, T.X, S. R. L.) and Atomic, Molecular, and Optical Sciences program (D.H. and M.H.G.). The authors would also like to acknowledge Paul Houston for his achievements in physical chemistry throughout the years and how his contributions to the field have helped shape the directions of future research.

\end{acknowledgement}

\bibliography{references}

\end{document}